\documentclass[aps,prl,showpacs,reprint]{revtex4-1}
\usepackage{dcolumn,graphicx,color}
\usepackage{amsfonts,amsmath,amssymb}
\usepackage{bm}% bold math
\newcommand{\bt}{\mathbf{t}}
\newcommand{\bb}{\mathbf{b}}
\newcommand{\bu}{\mathbf{u}}
\newcommand{\bn}{\mathbf{n}}
\newcommand{\bF}{\mathbf{F}}
\newcommand{\bM}{\mathbf{M}}

\begin{document}
\title{Leonardo's rule, self-similarity and wind-induced stresses in trees}
\author{Christophe Eloy}
\altaffiliation[Permanent address: ]{IRPHE, CNRS \& Aix-Marseille Universit\'e, Marseille, France}
\affiliation{Department of Mechanical and Aerospace Engineering, University of California San Diego, 9500 Gilman Drive, La Jolla CA 92093-0411, USA}
\date{\today}                                           
\begin{abstract}
Examining botanical trees, Leonardo da Vinci noted that the total cross-section of branches is conserved across branching nodes. In this Letter, it is proposed that this rule is a consequence of the tree skeleton having a self-similar structure and the branch diameters being adjusted to resist wind-induced loads. 
\end{abstract}
\maketitle

%%%%%%%%%%%%%%%%%%%%%%%%%%%%%%%%%%%%%%%%%%%%%%%%%%%%%%%%%%%%%%%%%%%%%%%%%%%%%%%%%%%%%%%%%%%%%%%%%%%%%%%%

Leonardo da Vinci observed in his notebooks that
``\emph{all the branches of a tree at every stage of its height when put together are equal in thickness to the trunk}'' \cite{Richter1998}, which means that when a mother branch of diameter $d$ splits into $N$ daughter branches of diameters $d_i$, the following relation holds on average
\begin{equation}
d^\Delta =\sum_{i=1}^N d_i^\Delta,
\end{equation}
where the Leonardo exponent is $\Delta =2$.  
\textcolor{black}{Surprisingly, there have been very few assessments of this rule, but the available data indicate that the Leonardo exponent is in the interval $1.8<\Delta<2.3$ for a large number of species \cite{Mandelbrot1983,Aratsu1998,Sone2009}. 
As a matter of fact, Leonardo's rule is so natural to the eye that it is routinely used in computer-generated trees \cite{Prusinkiewicz1990}.
Yet, alternative analyses of the branching geometry have been proposed based on analogies with river networks, bronchial trees, and arterial trees \cite{Zhi2001}.}

Two different models have been proposed to explain Leonardo's rule: the \emph{pipe model} \cite{Shinozaki1964}, which assumes that  trees are a collection of identical vascular vessels connecting the leaves to the roots, and the principle of \emph{elastic similarity} \cite{McMahon1976,Enquist2000}, which postulates that the deflection of branches under their weight is proportional to their length. However, none of these explanations are convincing. The first because the portion of a branch cross-section devoted to vascular transport (i.e. the sapwood) may be as low as $5\%$ in mature trees and it seems thus dubious that the whole tree architecture is governed by hydraulic constraints. \textmd{The second because the 
postulate behind elastic similarity is artificial, hard to relate to any adaptive advantage, and, furthermore, it seems unlikely that trees can respond to branch deflections.}
 
In this Letter, an alternative explanation is offered: Leonardo's rule is a consequence of trees being designed to resist wind-induced stresses. 
Plants are known to respond to dynamic loading for a long time, a phenomenon called \emph{thigmomorphogenesis} \cite{Niklas1992,Moulia2006a}. In that line of thinking, Metzger \cite{Metzger1893} proposed in the 19th century the \emph{constant-stress model}. This \textcolor{black}{model states that the trunk diameter} varies such that the bending stress due to wind remains constant \textcolor{black}{along the trunk length. The constant-stress model has been shown to agree with observations \cite{Morgan1994}, however, its implications} on the whole branching architecture has not yet been addressed \textcolor{black}{(except in the recent study of Lopez et al. \cite{Lopez2011})}. The other important point is that constant-stress might not be the best design since it  implies that breakage is more likely to occur in the trunk or in large branches where the presence of defects is more probable.

%%%%%%%%%%%%%%%%%%%%%%%%%%%%%%%%%%%%%%%%%%%%%%%%%%%%%%%%%%%%%%%%%%%%%%%%%%%%%%%%%%%%%%%%%%%%%%%%%%%%%%%%
% \section{Two models}
To address this problem, two equivalent analytical  models are  considered: one discrete, the \emph{fractal model}, and one continuous, the  \emph{beam model}, inspired from McMahon \& Kronauer \cite{McMahon1976} 
\textmd{with the difference that wind loads are considered instead of the weight.} 

%%%%%%%%%%%%%%%%%%%%%%%%%%%%%%%%%%%%%%%%%%%%%%%%%%%%%%%%%%%%%%%%%%%%%%%%%%%%%%%%%%%%%%%%%%%%%%%%%%%%%%%%
\begin{figure}[b] 
   \includegraphics[scale=0.4545]{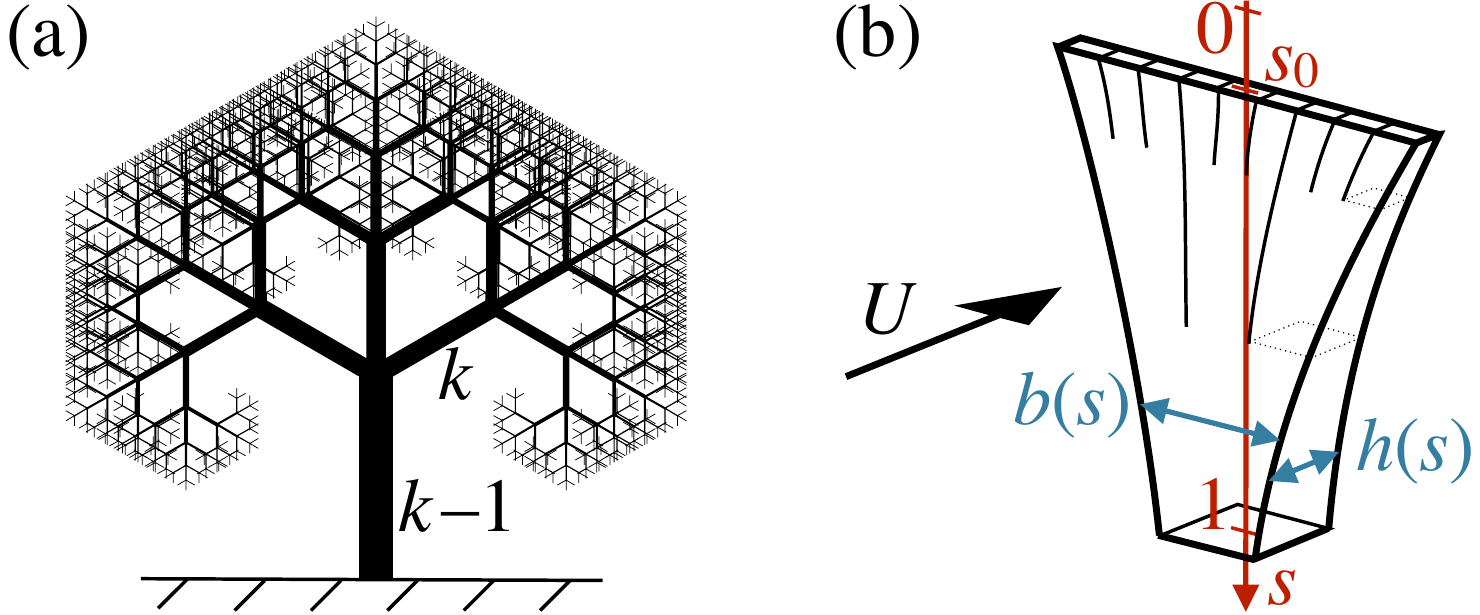} 
\caption{Two analytical models: (a) the fractal tree model; (b) the continuous tapered beam model \cite{McMahon1976}.}
\label{fig:sketch}
\end{figure}
%%%%%%%%%%%%%%%%%%%%%%%%%%%%%%%%%%%%%%%%%%%%%%%%%%%%%%%%%%%%%%%%%%%%%%%%%%%%%%%%%%%%%%%%%%%%%%%%%%%%%%%%

The fractal model (Fig.~\ref{fig:sketch}a) is constructed such that
\begin{equation}
\frac{l_k}{l_{k+1}}=N^\frac{1}{D}, \quad 
\frac{d_k}{d_{k+1}}=N^\frac{1}{\Delta},
\end{equation}
where $l_k$ and $d_k$ are the length and diameter of a branch at rank $k$ (with $1\le k\le K$), $N$ is the number of daughter branches at each branching node, $\Delta$ is Leonardo exponent, and $D$ is the fractal (Hausdorff) dimension of the tree skeleton \cite{Mandelbrot1983}. Here, the tree skeleton is supposed to be self-similar such that $D$ is uniform within the structure, but $\Delta$ can depend on $k$.

The fractal dimension $D$ has never been measured directly on real trees. However, the fractal dimension of the foliage surface has been measured to lie in the interval $2.2<D_\mathrm{fol.}<2.8$
\cite{Zeide1993} and, except for very particular architectures, it can be shown that $D=D_\mathrm{fol.}$. As already suggested by Mandelbrot \cite{Mandelbrot1983}, it can thus be safely assumed  that $2<D<3$ \footnote{Volume filling corresponds to $D=3$ and would maximize light interception. But trees tend to avoid having branches arbitrarily close to each other as it favor breakage when the tree sways. It results that $D<3$.}.

The  beam model (Fig.~\ref{fig:sketch}b) consists of a cantilevered beam whose width $b$ and thickness $h$ taper with the curvilinear coordinate $s$ (with $s_0\le s \le 1$).
These two models can be linked using the principle sketched in Fig.~\ref{fig:sketch}b. 
\textmd{It consists in cutting the beam to form branches of approximately square cross-sections. The beam thickness is then equivalent to the branch diameters, the ratio of width to thickness gives the number of branches, and $s$ corresponds to the branch lengths (because the distance between a branch and the tips is proportional to the branch length for infinite branching)}
\begin{equation}\label{eq:correspondence}
d_k \sim h, \quad
N^k \sim b/h\sim s^{-D}, \quad
l_k \sim s .
\end{equation}

%%%%%%%%%%%%%%%%%%%%%%%%%%%%%%%%%%%%%%%%%%%%%%%%%%%%%%%%%%%%%%%%%%%%%%%%%%%%%%%%%%%%%%%%%%%%%%%%%%%%%%%%
%\section{Loading and probability of fracture}

Consider now two different types of wind loads: either a continuous loading due to the wind on the branches with a force per unit length $q(s)\sim b$ or an end loading due to the wind in the leaves with a force $Q$ applied in $s_0$ equivalent to $q(s)=Q\delta(s-s_0)$. 
\textmd{Neglecting the wind incident angle and using the Euler-Bernoulli beam equation, the curvatures $k(s)$ resulting from the continuous load and the end load are found to scale respectively as}
\begin{equation}
k \sim \frac{q(s_0) s_0 s}{E I}, \quad
k \sim \frac{Q(s-s_0)}{E I},
\end{equation}
with $E$ the Young's modulus and $I\sim bh^3$ the moment of inertia. The expression for continuous loading is only valid for $s\gg s_0$  \footnote{It must also be assumed that the load diverges when $s\to 0$, i.e. if $b\sim s^\alpha$ near $s_0$, it requires $-2<\alpha<-1$. This latter assumption can be checked with (\ref{eq:model_finitesize}a,b) and is equivalent to $(4m-1)/(2m-1)<D<(7m-3)/(2m-1)$, i.e. $2.05<D<3.5$, for $m=10$.}.
Since the above scalings are equivalent at leading order, the analysis will be restricted to the case of end loading for simplicity. The maximum bending stress occurs at the beam surface and is $\sigma = Ekh/2$ such that 
\begin{equation}\label{eq:sigma}
\sigma \sim \frac{Q(s-s_0)}{bh^2}.
\end{equation}

The probability of fracture  at a given rank $k$ can be modeled by a Weibull distribution \cite{Bazant1998} to take into account size effects
\begin{equation}\label{eq:Weibull}
P_k = 1 -\exp\left[ -\frac{V_k}{V_0} \left(\frac{\sigma_k}{\sigma_0}\right)^m \right],
\end{equation}
where $V_k = N^k l_k \pi d_k^2 /4$ is the volume of all branches of rank $k$, $\sigma_0$ is the strength of the material, $V_0$ is an arbitrary volume taken to be $V_0=\pi l_1^3/4$ and $m$ is the Weibull's modulus (typically $5<m<20$ for wood \cite{Thelandersson2003}). 
\textmd{It can be shown that, for a given probability of fracture, the lightest design corresponds to the constant-stress model. However, since $V_k$ is  decreasing with $k$, this implies that the trunk and bottom branches are more likely to fail. As discussed in \cite{Lopez2011}, a better design is obtained when the probability of fracture $P_k$ is constant or increasing with $k$ such that the tree can regrow after a big storm. The equiprobability of fracture is expressed as}
\begin{equation}\label{eq:sigma_m}
\sigma^{-m} \sim V_k \sim hbs.
\end{equation}
\textmd{and it corresponds to $\sigma$ decreasing with $s$ as observed in trees \cite{Niklas2000a,Lopez2011}. 
When $P_k$ is increasing algebraically with $k$, the relation (\ref{eq:sigma_m}) still holds minor logarithmic correction.}

Now, using (\ref{eq:correspondence}), (\ref{eq:sigma}) and (\ref{eq:sigma_m}), the Leonardo  exponent and the diameter are found to depend on the fractal dimension $D$ and the Weibull's modulus $m$ 
\begin{subequations}\label{eq:model_finitesize}
\begin{eqnarray}
\Delta   &  =   & \frac{(3m-2)D(s-s_0)}{[(m-1)D+1](s-s_0)+ms},\\
h^{3m-2} & \sim & s^{(m-1)D+1} (s-s_0)^{m}.
\end{eqnarray}
\end{subequations}
In the case of infinite branching (i.e. $K=\infty$ or $s_0=0$), it gives $1.93<\Delta<2.21$ when $2<D<3$, for $m=10$. In other words, Leonardo's rule is recovered by assuming that the probability of fracture due to wind-induced stresses is constant. Note that, in (\ref{eq:model_finitesize}a,b), constant-stress corresponds to the limit $m\to\infty$. Note also that the number $N$ of daughter branches at each node does not affect the result.

%%%%%%%%%%%%%%%%%%%%%%%%%%%%%%%%%%%%%%%%%%%%%%%%%%%%%%%%%%%%%%%%%%%%%%%%%%%%%%%%%%%%%%%%%%%%%%%%%%%%%%%%%%
% Numerical model
%%%%%%%%%%%%%%%%%%%%%%%%%%%%%%%%%%%%%%%%%%%%%%%%%%%%%%%%%%%%%%%%%%%%%%%%%%%%%%%%%%%%%%%%%%%%%%%%%%%%%%%%
\begin{figure}[b] 
   \includegraphics[scale=0.227]{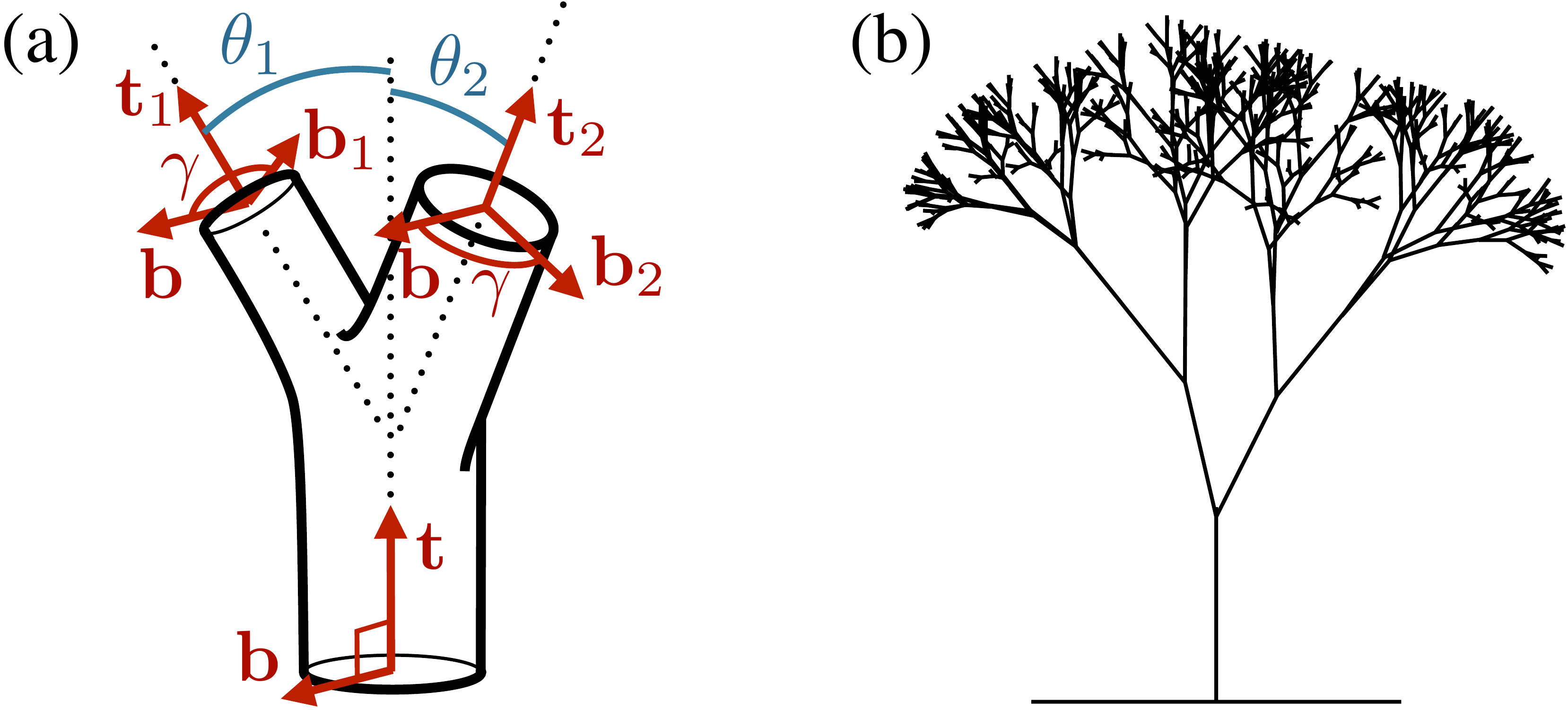} 
\caption{Numerical tree model: (a) Sketch of the angles and unit vectors at a branching node; (b) Example of tree skeleton for $\theta_1=-15^{\circ}$, $\theta_2=30^{\circ}$, $\gamma=120^{\circ}$, $r_1=r_2=0.75$, $p=1$, $K=10$, $D=2.41$ [as given by (\ref{eq:Dnumeric})].}
\label{fig:sketch2}
\end{figure}
%%%%%%%%%%%%%%%%%%%%%%%%%%%%%%%%%%%%%%%%%%%%%%%%%%%%%%%%%%%%%%%%%%%%%%%%%%%%%%%%%%%%%%%%%%%%%%%%%%%%%%%%

%%%%%%%%%%%%%%%%%%%%%%%%%%%%%%%%%%%%%%%%%%%%%%%%%%%%%%%%%%%%%%%%%%%%%%%%%%%%%%%%%%%%%%%%%%%%%%%%%%%%%%%%
\begin{figure*}[t] 
   \includegraphics[scale=.909]{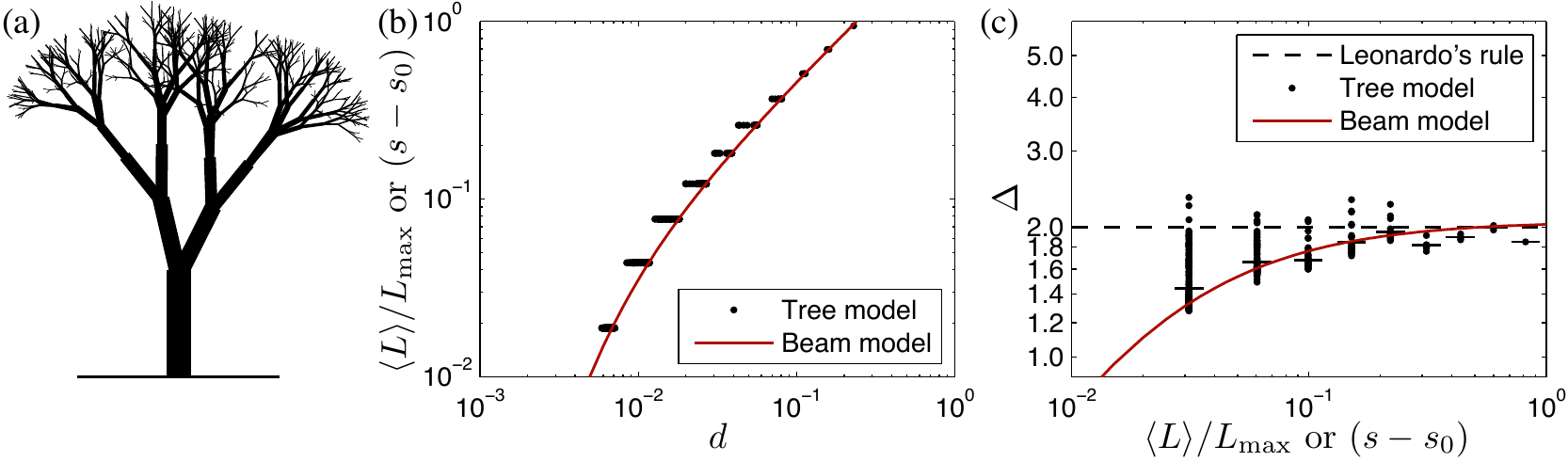} 
\caption{Deterministic tree: (a) Calculated branch diameters for the skeleton shown in Fig.~\ref{fig:sketch2}b; (b) Normalized average distance from the tips as a function of the diameter; (c) Calculated Leonardo exponent for each branching node (the horizontal bars show the mean value of $\Delta$ at each rank). In (a) and (b), the beam model corresponds to the relations (\ref{eq:model_finitesize}a,b). }
\label{fig:regular}
\end{figure*}
%%%%%%%%%%%%%%%%%%%%%%%%%%%%%%%%%%%%%%%%%%%%%%%%%%%%%%%%%%%%%%%%%%%%%%%%%%%%%%%%%%%%%%%%%%%%%%%%%%%%%%%%

To assess the robustness of these predictions when the asymmetry and stochasticity of branching, as well as the wind incident angle are taken into account, a three-dimensional numerical model has been developed. Following Niklas \& Kerchner \cite{Niklas1984}, a tree skeleton is recursively constructed as sketched in Fig.~\ref{fig:sketch2}a. 

Starting with a vertical trunk of length $l=l_\mathrm{trunk}=1$, parallel to the unit vector $\bt$ and normal to the unit vector $\bb$, two daughter branches of lengths $l_1=r_1 l$ and $l_2=r_2 l$ are constructed in the plane normal to $\bb$ such that their tangential unit vectors $\bt_1$ and $\bt_2$ are obtained by rotating $\bt$ with the angles $\theta_1$ and $\theta_2$ around $\bb$. The new normal vectors $\bb_1$ and $\bb_2$ defining the successive planes of branching are then obtain by rotating $\bb$ with an angle $\gamma$ around $\bt_1$ and $\bt_2$ respectively. This branching rule is recursively applied for $K$ ranks, with a probability of branching $p$, yielding a tree skeleton as examplified in Fig.~\ref{fig:sketch2}b. The architecture of this skeleton is parametrized by the six dimensionless quantities: $\theta_1$, $\theta_2$, $\gamma$, $r_1$, $r_2$ and $p$. This skeleton is self-similar with a fractal dimension 
\begin{equation}\label{eq:Dnumeric}
D=-\frac{\ln 2}{\ln r}, \quad\mbox{with } r=\frac{r_1+r_2}{2}p.
\end{equation}

Once this tree skeleton is constructed, the diameters of each branch can be calculated. Assuming that the wind velocity $U\bu$ (where $\bu$ is a unit vector) is parallel to the ground and uniform, the wind load on the leaves located at the tip of the terminal branch is 
\begin{equation}
\bF_{\mathrm{leaf}} = \tfrac{1}{2} \rho U^2 C_1 S_0 \bu ,
\end{equation}
where $\rho$ is the air density, $C_1$ is a drag coefficient which will be taken to be 1 without loss of generality and $S_0$ is the surface of the leaves assumed to be the square of the expected terminal branch length (i.e. $S_0=r^{2(K-1)}l_\mathrm{trunk}^2$).

In addition, the wind exerts also a force on each branch such that, if $\bn$ is the unit vector normal to both the wind and the branch (i.e. $\bn= \bt\times\bu/\|\bt\times\bu \|$), the force exerted on each branch is 
\begin{equation}
\bF_{\mathrm{branch}} = \tfrac{1}{2} \rho U^2 C_2 d l \|\bt\times\bu \|^2 (\bn\times\bt) ,
\end{equation}
where $C_2$ is another drag coefficient taken to be 1, $d$ and $l$ are the diameter and the length of the branch, and $\|\bt\times\bu \|^2$ is the square of the incident angle cosine. This force is applied on the branch center of mass such that its moment at the base of the branch is simply $\bM_{\mathrm{branch}} = \frac{1}{2}l\, \bt\times\bF_{\mathrm{branch}}$.

Now each branch transmits the forces and moments applied at its extremity (either originating from upper branches or by leaves) such that, if $\bF_{\mathrm{top}}$ and $\bM_{\mathrm{top}}$ are the sum of forces and moments at a branch end, the force and moment at the branch base are
\begin{subequations}
\begin{eqnarray}
\bF_{\mathrm{base}} & = & \bF_{\mathrm{branch}} + \bF_{\mathrm{top}},\\
\bM_{\mathrm{base}} & = & \bM_{\mathrm{branch}} + \bM_{\mathrm{top}} + l\,\bt\times\bF_{\mathrm{top}}.
\end{eqnarray}
\end{subequations}
The moment at the base $\bM_{\mathrm{base}}$ has two components: a bending moment of intensity $M_{\mathrm{bend}}= \| \bM_{\mathrm{base}} \times \bt \|$ and a torsional moment of intensity $M_{\mathrm{twist}}= \| \bM_{\mathrm{base}} \cdot \bt \|$. The corresponding maximal bending (tensile and compressive) and shear stresses are $\sigma_{\mathrm{bend}} = \frac{32}{\pi} M_{\mathrm{bend}} / d^3$ and $\sigma_{\mathrm{shear}} = \frac{32}{\pi} M_{\mathrm{twist}} / d^3$.

%%%%%%%%%%%%%%%%%%%%%%%%%%%%%%%%%%%%%%%%%%%%%%%%%%%%%%%%%%%%%%%%%%%%%%%%%%%%%%%%%%%%%%%%%%%%%%%%%%%%%%%%
%\section{Results}

Assuming that there is a uniform probability  of fracture $(1-\mathrm{e}^{-1})$ for every rank as given by (\ref{eq:Weibull}), the diameter of each branch can  be  calculated recursively, starting from the tips and ending with the trunk. In doing so, resistance to bending and twisting has been considered and the wind direction has been assumed to vary with increments of $45^{\circ}$. 
In this calculation, the Cauchy number, $C_Y=\rho U^2 / \sigma_{0,\mathrm{bend}}$, appears as the dimensionless parameter which sets the scaling of branch diameters (but not their relative values) such that $d\sim C_Y^{m/(3m-2)} l_\mathrm{trunk}$. It has been taken to be $C_Y=10^{-4}$ which corresponds roughly to $U=40\,$m$\,$s$^{-1}$ and $\sigma_{0,\mathrm{bend}}=20\,$MPa \cite{Thelandersson2003}.
The other important dimensionless numbers are the relative surface of leaves $S_0$ (which sets the total height of the tree assuming that leaves have always the same dimension whatever the size of the tree), and the ratio of bending to shear strength $\sigma_{0,\mathrm{bend}} / \sigma_{0,\mathrm{shear}}$ taken to be equal to 5 as it is generally observed for wood \cite{Thelandersson2003}.

%%%%%%%%%%%%%%%%%%%%%%%%%%%%%%%%%%%%%%%%%%%%%%%%%%%%%%%%%%%%%%%%%%%%%%%%%%%%%%%%%%%%%%%%%%%%%%%%%%%%%%%%
\begin{figure*}[t]
   \includegraphics[scale=.909]{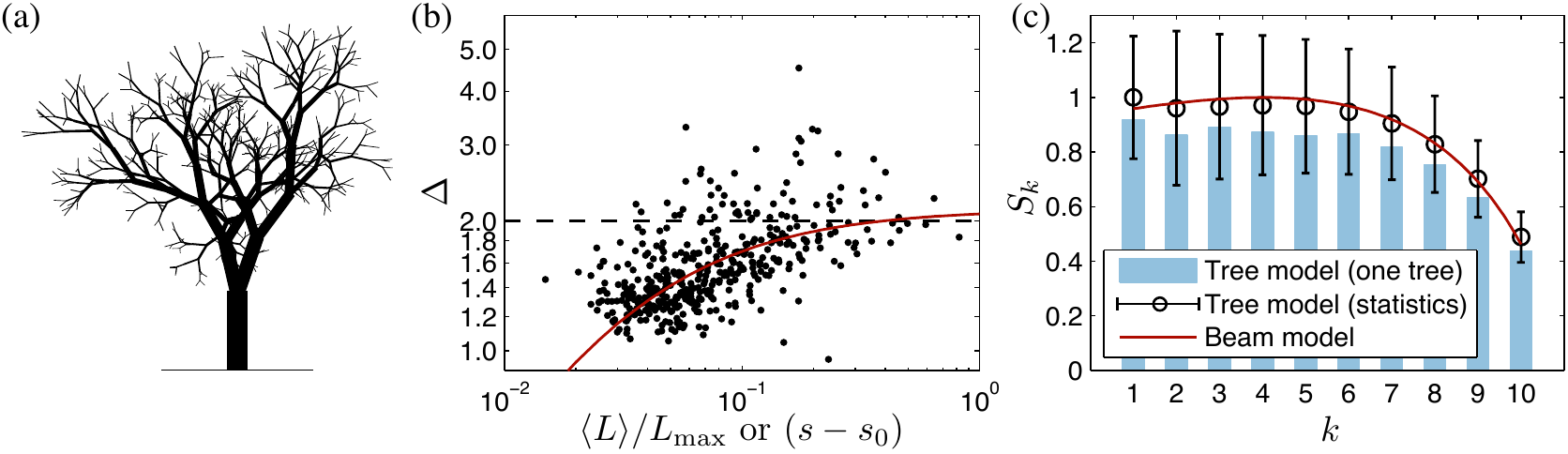} 
\caption{Stochastic tree: (a) Silhouette of a tree with parameters: $\bar{\theta}_1=-20^{\circ}$, $\bar{\theta}_2=40^{\circ}$, $\bar{\gamma}=100^{\circ}$, $\bar{r}_1=0.75$, $\bar{r}_2=0.85$, $p=0.98$, $K=10$, $D=2.85$; (b) Corresponding Leonardo exponent (same legend as in Fig.~\ref{fig:regular}c); (c) Total cross-sectional area at every rank $k$ normalized by its maximum. The bars correspond to the tree depicted in (a), the curve is the prediction from the beam model and the symbols with the error bars correspond to the average and standard deviation on 1000 realizations where the mean branching angles are randomly taken in the intervals $-60^{\circ}<\bar{\theta}_1<0$, $0<\bar{\theta}_2<60^{\circ}$, $-180^{\circ}<\bar{\gamma}<180^{\circ}$.}
\label{fig:random}
\end{figure*}
%%%%%%%%%%%%%%%%%%%%%%%%%%%%%%%%%%%%%%%%%%%%%%%%%%%%%%%%%%%%%%%%%%%%%%%%%%%%%%%%%%%%%%%%%%%%%%%%%%%%%%%%

The result of such a calculation is shown in Fig.~\ref{fig:regular} for the deterministic skeleton pictured in Fig.~\ref{fig:sketch2}b. To compare these results with the theoretical predictions, the ratio $\langle L \rangle / L_\mathrm{max}$ is used \cite{McMahon1976}, where  $\langle L \rangle$ is the average distance from the branch tips considering all possible paths and $L_\mathrm{max}=1/(1-r)$  is the mean ground-to-tips distance for an infinitely branching tree. The ratio $\langle L \rangle / L_\mathrm{max}$ is equivalent to $(s-s_0)$ for the beam model.  

As seen in Fig.~\ref{fig:regular}, the  beam model accurately predicts the branch diameters and the Leonardo exponent. It means that the wind incident angle and the geometric details of  branching do not affect these scalings. Note that, because of the finite number of recursions, the slope in  Fig.~\ref{fig:regular}b is not constant as already observed in \cite{McMahon1976}. % Note also that for the particular set of parameter chosen for Fig.~\ref{fig:regular}, the torsional moments can be neglected since they play no role. 

In Fig.~\ref{fig:random}, the same simulation is run except that, at each branching node, the angles $\theta_1$, $\theta_2$, $\gamma$ are randomly chosen with a normal distribution of means $\bar{\theta}_1$, $\bar{\theta}_2$, $\bar{\gamma}$ and standard deviation of $10^{\circ}$. Same is done for $r_1$ and $r_2$ of means $\bar{r}_1$ and $\bar{r}_2$ and standard deviation of $0.1$. It results that the Leonardo exponent is more scattered but the beam model still predicts it correctly (Fig.~\ref{fig:random}b). 

Figure~\ref{fig:random}c shows directly the prediction of Leonardo's rule by comparing the total cross-sectional areas at every rank. Depending on the particular angles defining the branching rules, this surface can vary with a standard deviation of about $20\%$  but its mean remains roughly constant for all ranks except the last three \footnote{For the smallest branches, the vascular function cannot be neglected anymore. The present prediction of their diameters can be viewed as a lower branch.}. Remarkably, the variation of this mean is accurately predicted by the continuous beam model.

%%%%%%%%%%%%%%%%%%%%%%%%%%%%%%%%%%%%%%%%%%%%%%%%%%%%%%%%%%%%%%%%%%%%%%%%%%%%%%%%%%%%%%%%%%%%%%%%%%%%%%%%
%\section{Conclusion}

In summary, it has been shown that the best design to resist wind-induced fracture in self-similar trees naturally yields Leonardo's rule.  
The only requirement is that trees adapt their local growth to wind loads, a well-known phenomenon called thigmomorphogenesis whose mechanism at the cell level is still largely unknown. 
Here, the relevant property of wind loads is their divergence towards the branch tips, either because of the leaves or because the surface exposed to wind diverges. Thus the static loads due to the weight of fruits, snow, or ice would give similar results. 

The validity of the present model could be assessed experimentally by examining how the branch diameters depend on the wind peak velocity. It is predicted here that the following relation holds: $d\sim C_Y^{m/(3m-2)} l_\mathrm{trunk}$, where $C_Y$ is proportional to $U^2$.
Another assessment would be calculate, from a real tree skeleton, the expected branch diameters and compare them to the measured ones. 
In this Letter, aeroelastic reconfiguration \cite{Vogel2009}, branch weight \footnote{From the tree of Fig.~\ref{fig:random}, it has been calculated that bending moments due to the branch weights are negligible when $\rho_{\mathrm{tree}} g \langle L \rangle / \rho U^2 \lesssim 3\, C_Y^{m/(2-3m)}$. Using a wood density of $\rho_{\mathrm{tree}} =800\,$kg$\,$m$^{-3}$, it yields $\langle L \rangle \lesssim 20\,$m. For trees taller than that gravity should be taken into account, at least for the first branches above the trunk.}, and non-uniform wind profiles \cite{Niklas2000a}, have been neglected for the sake of simplicity. 
It has also been assumed that the tree skeleton is fractal, with a fractal dimension $2<D<3$. Yet, the way $D$ and  other features of the tree skeleton depend on the wind, and on the environment in general, remains to be explored.

\begin{acknowledgments}
I warmly thank Emmanuel de Langre for having inspired this work and Eric Lauga for stimulating discussions. This study was supported by the European Union through the fellowship PIOF-GA-2009-252542.
\end{acknowledgments}

\end{document}